\documentclass[%
 reprint,
 amsmath,amssymb,
 aps, nofootinbib,superscriptaddress
]{revtex4-2}
\usepackage{amsmath}
\usepackage{amssymb}
\usepackage{amsfonts}
\usepackage[utf8]{inputenc}
\usepackage[T1]{fontenc}
\usepackage{mathrsfs}
\usepackage{anyfontsize}

\usepackage[linktocpage,breaklinks]{hyperref}
\usepackage[usenames,dvipsnames]{xcolor}
\usepackage{txfonts}
\usepackage{tensor}
\usepackage{bm}

\usepackage{graphicx}
\usepackage{epsfig}
\usepackage{epstopdf}

\usepackage{natbib}
\usepackage{hyperref}

\hypersetup{colorlinks=true, citecolor=Purple, linkcolor=Purple,
	urlcolor=Purple}
\usepackage{graphicx}
\usepackage{dcolumn}
\usepackage{bm}

\begin{document}

\preprint{APS/123-QED}

\title{Quasibound state reminiscent in de Sitter black holes: Quasinormal modes and the decay of massive fields}

\author{Mateus Malato Corrêa}
 \email{malato.mateus@gmail.com}\affiliation{
 Programa de P\'{o}s-Gradua\c{c}\~{a}o em F\'{i}sica, Universidade Federal do Par\'{a}, 66075-110, Bel\'{e}m, PA, Brazil
}
\author{Caio F. B. Macedo}
 \email{caiomacedo@ufpa.br}
\affiliation{
 Programa de P\'{o}s-Gradua\c{c}\~{a}o em F\'{i}sica, Universidade Federal do Par\'{a}, 66075-110, Bel\'{e}m, PA, Brazil
}
\affiliation{Faculdade de F\'{i}sica, Campus Salin\'{o}polis, Universidade Federal do Par\'{a}, 68721-000, Salin\'{o}polis, Par\'{a}, Brazil
}
\author{João Luís Rosa}
\email{joaoluis92@gmail.com}
\affiliation{Institute of Theoretical Physics and Astrophysics, University of Gda\'{n}sk, Jana Ba\.{z}y\'{n}skiego 8, 80-309 Gda\'{n}sk, Poland}
\affiliation{Institute of Physics, University of Tartu, W. Ostwaldi 1, 50411 Tartu, Estonia}

\date{\today}

\begin{abstract}
Massive perturbations in asymptotic flat black holes leave a distinct signature in their late-time evolution ``tail'': an oscillatory behavior modulated by the Compton wavelength of the field, which can be associated with the so-called quasibound state spectrum. In asymptotically de Sitter spacetimes, however, the massive perturbations always leak to the cosmological horizon, which indicates the absence of a quasibound part of the spectrum. In this work, we show that an additional mode exists in asymptotically de Sitter black holes that produces an imprint similar to that of the quasibound states in the late-time behavior of massive scalar perturbations. If the Compton wavelength is larger than a certain critical value (which depends on the cosmological constant), the oscillatory behavior of the tail turns into an exponential decay due to the fact that the de Sitter mode is purely imaginary. Even for black holes with typical length scales small in comparison to the size of the cosmological horizon, the late-time tail behavior of the massive perturbations is modified as compared to the usual $t^{-5/6}$ for Schwarzschild black holes, thus being a distinctive feature induced by the presence of a cosmological constant.
\end{abstract}
\maketitle

\section{\label{sec:Introduction}Introduction}

Black holes (BHs) are the standard objects with strong gravitational fields. Their signature comes from many astrophysical channels, such as electromagnetic and gravitational wave observations~\cite{2016:LIGOVIRGO,2017:LIGOVIRGO,2019:EHTcollab,EventHorizonTelescope:2021bee,2022:EHTCollab,GRAVITY:2020lpa,GRAVITY:2020gka,GRAVITY:2023avo}. The dynamics of BHs in General Relativity (GR) are fairly understood, covering perturbative post-Newtonian methods to full-blown numerical relativity. In particular, the quasinormal spectrum of BHs is crucial to understanding how these objects relax to a final state of equilibrium~\cite{1999:Nollert,2009:BertiCardosoStarinets,2011:KonoplyaZhidenko}.

\begin{figure}[h]
    \includegraphics[width=1\linewidth]{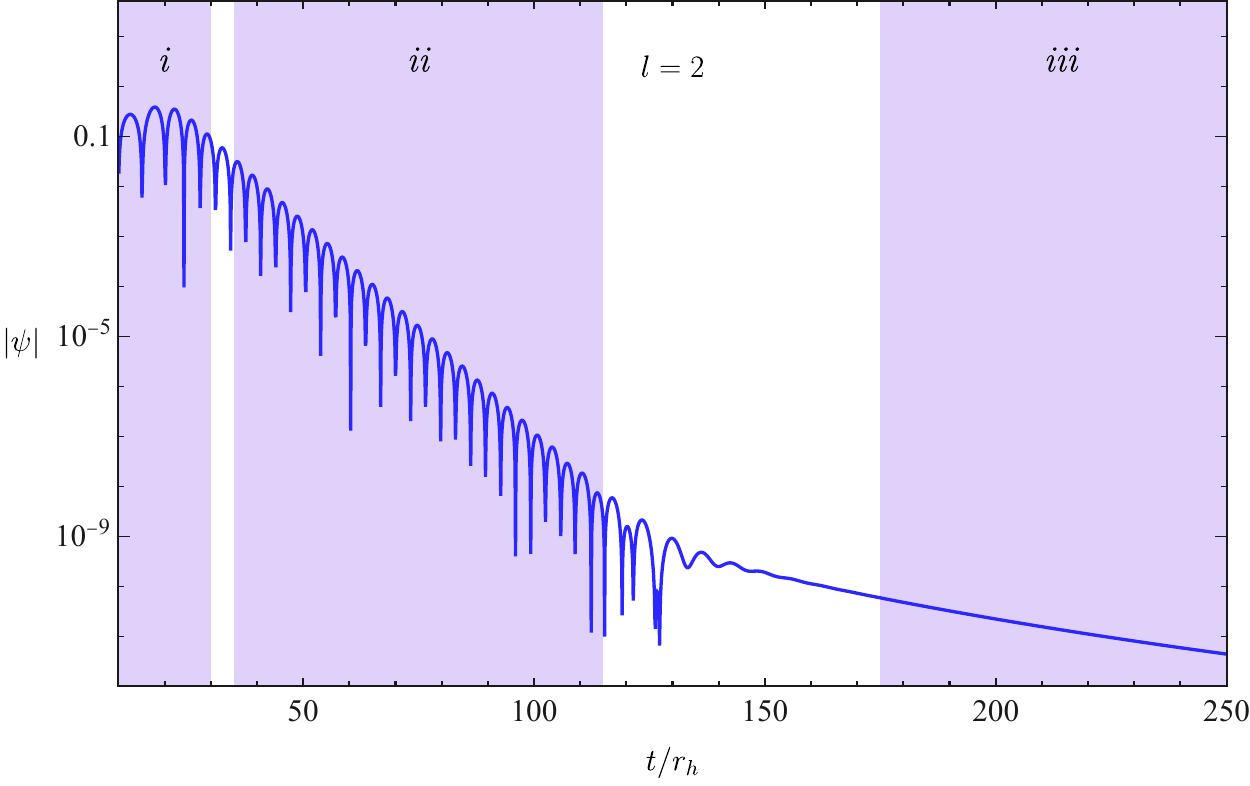}
   \caption{Time evolution of a massless scalar field in Schwarzschild spacetime, highlighting each phase: \textit{i}) the initial prompt, \textit{ii}) the ringdown phase, and \textit{iii}) the late time.}
    \label{FIG:Schwarl2}
\end{figure}

Perturbations of BHs and the study of quasinormal modes (QNM) have been the subject of intensive study in the literature. The seminal studies of the stability of the Schwarzschild spacetime~\cite{1957:ReggeWheeler,1970:Zerilli,1970:Vishveshwara} paved the way to analyze the mode stability in different scenarios, Reissner-Nordstr\"om~\cite{1988:kokkotasSchutz,1990:Leaver} and Kerr~\cite{1972:Teukolsky} being the natural extension, but also testing different objects within and beyond GR~\cite{Molina:2010fb,Blazquez-Salcedo:2016enn,Macedo:2018txb,Berti:2018vdi,Cardoso:2019mqo,McManus:2019ulj,2023:Zeus}. The standard picture of the evolution of initial perturbations is \textit{i)} an initial prompt that depends on the initial conditions; \textit{ii)} the ringdown phase, in which the real part of the modes dictate the oscillatory pattern while the imaginary part the decaying of the signal; and \textit{iii)} finally in the late time, where we have no oscillations and decay of the field, usually described by a power law behavior. These features are illustrated in Fig.~\ref{FIG:Schwarl2} where we show the evolution of a scalar Gaussian wave packet in Schwarzschild spacetime. 

The standard evolution profile depicted above depends on the type of perturbation induced. For instance, scalar fields in asymptotic flat black hole spacetimes evolve differently whether they are massive or charged~\cite{2004:BurkoANDKhanna,Koyama:2000hj,Koyama:2001ee,Koyama:2001qw}.
Some of these properties can efficiently confine the perturbations, decreasing their leaking to infinity. As a result, we have a modification of the late-time behavior due to long-lived modes \cite{2005:KonoplyaZidenko,2004:BurkoANDKhanna,2004:OhashiSakagami,2016:Hod}. These modes may be perceived as quasiresonant modes or quasibound states (QBS), akin to the hydrogen spectrum, but with an imaginary part due to the dissipative nature of the event horizon. These types of modes have been analyzed in different spacetimes \cite{Berti:2019wnn,Cardoso:2015zqa,2004:OhashiSakagami,2002:konoplya,2007:SamDolan,1980:Detweiler,1970:DamourDaruelleRuffini}.
Additionally, whenever superradiance is present, this trapping can trigger instabilities in the spacetime \cite{2007:SamDolan,2004:CardosoDiasLemosYoshida,Rosa:2020uoi,1970:DamourDaruelleRuffini}, and can be considered as a natural ``mirror'' for the black hole bomb phenomena \cite{1972:PressTeulkolski,2020:BritoCardosoPani}.  As we show in what follows, this effect is nonexistent when asymptotic de Sitter spacetimes are considered, due to the existence of a cosmological horizon.

Asymptotically de Sitter spacetimes are solutions with a positive cosmological constant. Studies on massless scalar perturbations in de Sitter spacetimes, without a black hole, found purely imaginary quasinormal frequencies (QNF) \cite{2006:Lopez-Ortega2,2012:Lopez-Ortega}. Considering a massive field instead, one can modify the evolution of the perturbation leading to the presence of an oscillation, i.e., it can modify the quasinormal spectrum by introducing a real part to the frequencies \cite{2006:Lopez-Ortega,2012:Lopez-Ortega}. 

The Schwarzschild-de Sitter (SdS) spacetime inherits some properties from both the black hole and cosmological sides. It has been shown that QNMs in SdS exist in two families, one related to the black hole and the other to the de Sitter asymptotic behavior, which are known as photon sphere (PS) modes and de Sitter (dS) modes, respectively \cite{2020:Aragon,2012:Lopez-Ortega,2018:CardosoCostaDestounisJanses, 2022:KonoplyaZidenko}. The PS modes reduce to the usual Schwarzshild BH spectrum when we take the limit of the vanishing cosmological constant and, in the evolution of initial perturbations, they dictate the ringdown. In contrast, the de Sitter modes are purely imaginary and retrieve the de Sitter spectrum when the BH mass goes to zero, and are better seen in the late time tail \cite{2022:KonoplyaZidenko}. 

The QNF spectrum has been calculated for massive scalar fields in SdS spacetime in \cite{2020:Aragon,2017:ToshmatovStuchlik}. In \cite{2020:Aragon} the authors found that the mass can remove the anomalous decay for the PS modes (where modes with larger angular momentum numbers are more stable than the lower modes), and introduce a real part in the dS modes, inducing oscillations. Increasing the mass can also lead to a smaller imaginary part, similar to the asymptotically flat case \cite{2017:ToshmatovStuchlik,2005:KonoplyaZidenko}. Additionally, it has been shown that massive charged fields in Reissner-Nordström-de Sitter may present some long-lived ringing, and for massless perturbations the dS modes can be extracted from the late-time tails~\cite{2022:KonoplyaZidenko,2014:Zhu}. From these, we expect to relate the late-time tails and the dS modes for different values for the mass of the field.

As previously stated, the cosmological horizon prevents QBS from forming in SdS BHs. Nonetheless, since many of the features of the spacetime are shared with their asymptotically flat counterparts, we should expect that at some point we can recover the same signatures as the cosmological constant tends to zero. Moreover, since the effective potential changes asymptotically, the universal late-time behavior for massive fields, $t^{-5/6}$, should change when the cosmological constant is considered. These issues are addressed in what follows.

In this work we analyze the spectrum and the time evolution of perturbations of the massive scalar field in the SdS spacetime. In Sec.~\ref{sec:SF in SdS} we lay out the scalar field equation and its decomposition. In Sec.~\ref{sec:qnms} we outline the methods employed to compute the modes spectra. In Sec.~\ref{sec:results} we provide our numerical results, confirming that QBS are nonexistent even in the limit $\Lambda\to0$ ($r_c\to\infty$), with $\Lambda$ being the cosmological constant ($r_c$ the cosmological horizon). We also investigate the influence of the mass in the late-time tail and determine whether the oscillations due to the real part in the dS modes can be seen as reminiscent quasibound states from the Schwarzschild case by extracting the frequencies through the Prony method. Finally, in Sec.~\ref{sec:conclusion} we present our final remarks. Throughout this work we use the metric signature $(-,+,+,+)$ and a geometrized unit system for which $G=c=1$.

\section{Scalar field in Schwarzschild-de Sitter spacetime}\label{sec:SF in SdS}
The Schwarzschild-de Sitter (SdS) spacetime is a static and spherically symmetric BH solution with a cosmological horizon. The line element that describes such a spacetime in the standard Schwarzschild-like coordinates ($t,r,\theta,\phi$) is
\begin{equation}\label{EQ:SphericallySymmetric}
	ds^{2}=-f(r) dt^{2}+\dfrac{1}{f(r)}dr^{2}+r^{2}(d\theta^{2}+\sin^{2}\theta \,d\phi^{2}),
\end{equation}
where we have defined
\begin{equation}\label{EQ:lapsefunction}
    f(r)=1-\frac{2M}{r}-\frac{\Lambda}{3}r^{2} ,
\end{equation}
with $M$ being the mass of the BH and $\Lambda$ the cosmological constant. The radii of the event horizon $r_h$ and the cosmological horizon $r_{c}$ can be extracted by analyzing the roots of $f(r)$, i.e., $f\left(r_h\right)=f\left(r_c\right)=0$. This analysis allows one to rewrite the mass $M$ and the cosmological constant $\Lambda$ in terms of the radii of the horizons as:
    \begin{equation}\label{EQ:Mandalambdaintermsrhrc}
		M= \frac{r_{c}r_{h}(r_{c}+r_{h})}{2(r^{2}_{c}+r_{c}r_{h}+r^{2}_{h})}, \hspace{0.5 cm} \Lambda=\frac{3}{r^{2}_{c}+r_{c}r_{h}+r^{2}_{h}}.
    \end{equation} 
Following the results above, one can rewrite the lapse function $f(r)$ in a more suitable way as
\begin{equation}
    f(r)=\frac{\Lambda}{3}\frac{(r_{c}-r)(r-r_{h})(r+r_{c}+r_{h})}{r},
\end{equation}
which we shall use to parametrize the computations that follow. Note that $r_h\leq r_c$, and also that $r_c\to \infty$ and $r_h\to 2M$ in the limit $\Lambda\to0$, thus recovering the Schwarzschild spacetime.

\begin{figure*}[th]
        \includegraphics[width=.48\linewidth]{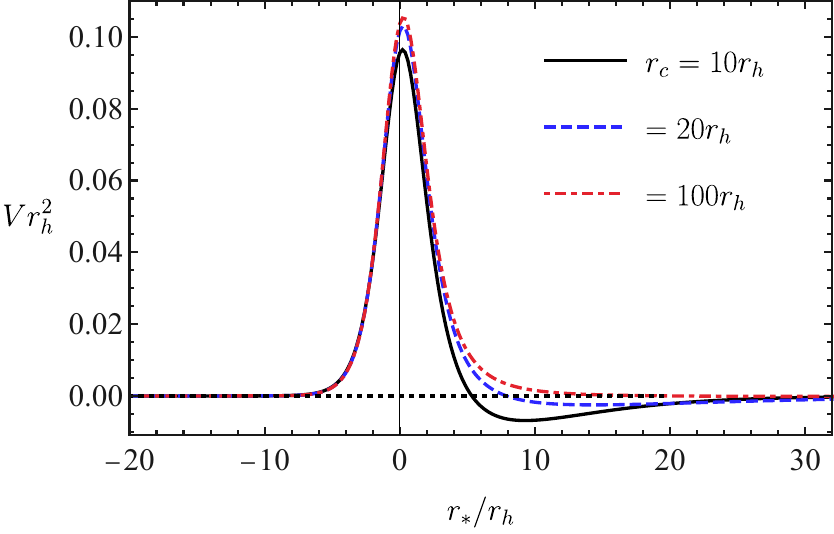}\qquad
        \includegraphics[width=.48\linewidth]{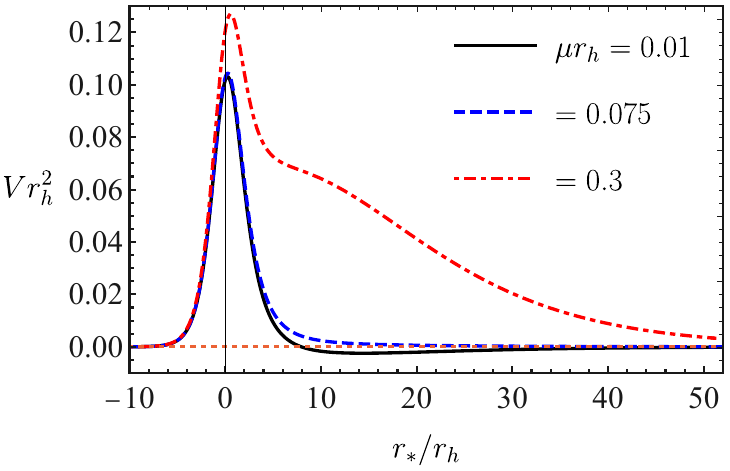}
        \caption{The effective potential $V$ as a function of $r_{*}$ with $l=0$, for a constant $\mu=0$ and different values of $r_{c}$  (left panel) and for a constant $r_c=20 r_h$ and different values of the field mass $\mu$ (right panel).}
        \label{FIG:PotentialVARlambdaMu}
    \end{figure*}

Consider now a massive scalar field $\Psi$. The equation of motion for such a scalar field is given by the Klein-Gordon equation $\left(\Box-\mu^2\right)\Psi=0$, where $\Box=g^{\alpha\beta}\nabla_\alpha\nabla_\beta$ is the d'Alembert operator, with $\nabla_\alpha$ denoting the covariant derivatives written in terms of the metric $g_{\alpha\beta}$, and $\mu$ is the mass of the scalar field. This equation can be written in the form
\begin{equation}\label{EQ: massivescalarfield DE}
     \dfrac{1}{(-g)^{1/2}}\partial_{\alpha}\left[(-g)^{1/2}g^{\alpha\beta}\partial_{\beta}\Psi\right]-\mu^{2}\Psi=0,
\end{equation}
where $g$ is the metric determinant, and $g^{\alpha\beta}$ the contravariant metric. Given the spherical symmetry of the spacetime under analysis, it is convenient to introduce the following decomposition of the scalar field as
    \begin{equation}\label{EQ:Psi}
            \Psi(t,r,\theta,\phi)=\dfrac{\psi(t,r)}{r} Y_{lm}(\theta,\phi)
    \end{equation}
where $Y_{lm}(\theta,\phi)$ denote the spherical harmonics, with $l$ and $m$ the angular momentum and magnetic numbers, respectively, and $\psi\left(t,r\right)$ is an angle-independent wave function. Furthermore, introducing a redefinition of the radial coordinate into the so-called tortoise coordinate $r_*$ defined by
    \begin{equation}\label{EQ:tortoise}
        dr_{*}=\frac{dr}{f(r)},
    \end{equation}
the Klein-Gordon equation given in Eq. \eqref{EQ: massivescalarfield DE} takes the form of a Schrödinger wavelike equation as
     \begin{equation}\label{EQ:diffpsi}
    \frac{\partial^2\psi(t,r_*)}{\partial r_*^2}-\frac{\partial^2\psi(t,r_*)}{\partial t^2}-V(r)\psi(t,r_{*})=0,
    \end{equation}
where the effective potential $V$ is defined as
    \begin{equation}\label{EQ:potential}
        V(r)=\dfrac{f(r)}{r^{2}}\left[l(l+1)+r^{2}\mu^{2}+r f'(r)\right].
    \end{equation}
We can further decompose the time dependence in terms of the field angular frequency $\omega$ as $\psi(t,r_*)\equiv \psi(r_*)e^{-i\omega t}$, the so-called harmonic ansatz, leading to
\begin{equation}\label{EQ:diffpsiR}
\frac{d^2\psi(r_*)}{d r_*^2}+\left[\omega^2-V(r)\right]\psi(r_{*})=0.
\end{equation}
In Fig.~\ref{FIG:PotentialVARlambdaMu} we plot the behavior of the effective potential $V$ for different values of the mass of the field $\mu$ and cosmological horizon $r_c$. Two distinctive features induced by the cosmological constant are \textit{i)} the existence of a region for which the potential is negative (which happens solely for $l=0$), and \textit{ii)} the vanishing of the potential as $r_*\to\infty$, even for the massive case. As such, instead of having an exponential suppression of the field, the solutions behave asymptotically as waves, in both boundaries. As the cosmological constant increases for a constant mass of the field, the maximum value of the potential decreases, the region in which the effective potential is negative broadens, and the minimum value of the potential decreases. On the other hand, the mass of the field induces a different behavior in the potential. As the mass increases for a constant value of the cosmological constant, the whole effective potential increases in the region $r_*>0$, eventually compensating for the negative contribution of the cosmological constant. In the region $r_*<0$, the potential resembles the Schwarzschild BH case.

\section{Quasinormal Modes and evolution} \label{sec:qnms}

Quasinormal mode frequencies are found by requiring the solutions of the wave equation to satisfy a set of physically motivated boundary conditions. 
Usually, in BH spacetimes, this implies that the scalar field must behave asymptotically as follows
\begin{equation}\label{EQ:BoundaryPsiSdS}
	\psi(r_{*})\sim\begin{cases}
	e^{G(\omega)\,r_{*}}, \text{ for } r_*\rightarrow -\infty \\
	r^{\alpha}e^{F(\omega)\,r_{*}}, \text{ for } r_*\rightarrow \infty,
	\end{cases}
    \end{equation} 
where the functions $(F,G)$ depend on the asymptotic forms of the spacetime, i.e., the behavior of the effective potential at the boundaries. The $\alpha$ parameter depends on characteristics of the physical system~\cite{2007:SamDolan,Richartz:2014jla}. For massive fields in Schwarzschild BHs, for instance, we have that $G(\omega)=-i\omega$ and $F(\omega)=\pm \sqrt{\mu^2-\omega^2}$. Note that at infinity, in this case, when $Re[F(\omega)]<0$ the solution is exponentially damped, corresponding to the QBS. On the other hand, QNMs are found by considering $Re[F(\omega)]>0$ and the solutions are wavelike at infinity, exponentially growing. \cite{2007:SamDolan}.
\footnote{Note that in other scenarios the boundary condition at the horizon might change, as in the case of charged scalar fields in charged or rotating BHs~\cite{2007:SamDolan,Richartz:2014jla}.} 
Both quasinormal and quasibound spectra coexist, having different imprints in the evolution of the fields. For instance, the quasibound spectrum is relevant in computing the superradiant instabilities in BH spacetimes (see Ref.~\cite{2020:BritoCardosoPani} and the references therein). In SdS spacetime, since the effective potential vanishes at both boundaries, one can write $G(\omega)=-i\omega$ and $F(\omega)=i\omega$, and therefore we have a wavelike behavior in both boundaries independently of the mass of the field, i.e., the quasibound spectrum is absent. One thus obtains
    \begin{equation}\label{EQ:BoundaryPsiSdS2}
		\psi(r_{*})\sim\begin{cases}
			e^{- i\,\omega\,r_{*}}, \text{ for } r_*\to-\infty~~ (r\to r_{h}) \\
			e^{i\,\omega\,r_{*}}, \text{ for } r_*\to\infty~~ (r\to r_{c}).
		\end{cases}
\end{equation}
In what follows, we apply three different methods to compute the QNF of SdS BHs. 

\subsection{Continued fraction method}\label{SUBSEC:COntFrac}

In this section, we implement the continued fraction method, also known as the Leaver's method, to extract the QNFs of massive scalar fields in SdS and Schwarzschild spacetimes, separately. Such a distinction is necessary since the results for the Schwarzschild spacetime cannot be directly obtained as a limit of those in SdS spacetimes when $\Lambda\to 0$ due to the fundamental difference in the boundary conditions stated previously, as we clarify in what follows.

\subsubsection{Schwarzschild-de Sitter Spacetime}
Let us now expose how we can implement the boundary conditions in Eq. \eqref{EQ:BoundaryPsiSdS2} to find the quasinormal spectrum of massive scalar fields in SdS. First, we note that for $r\to r_{h}$ we have
	\begin{equation}\label{EQ:EXPtortoise}
		e^{-i \omega r_{*}}=\left(\frac{1}{r}-\frac{1}{r_{h}}\right)^{\rho_{h}}\left(\frac{1}{r_{c}}-\frac{1}{r}\right)^{- \rho_{c}}\left(\frac{1}{r}+\frac{1}{r_{c}+r_{h}}\right)^{\rho_{h}-\rho_{c}},
	\end{equation}
	with
     \begin{equation}
     \begin{aligned}
	\rho_{h}=\frac{i\,\omega}{2\,M\left(\frac{1}{r_{c}}-\frac{1}{r_{h}}\right)\left(\frac{1}{r_{h}}+\frac{1}{r_{h}+r_{c}}\right)},\\ 
        \rho_{c}=\frac{i\,\omega}{2\,M\left(\frac{1}{r_{c}}-\frac{1}{r_{h}}\right)\left(\frac{1}{r_{c}}+\frac{1}{r_{h}+r_{c}}\right)}.
    \end{aligned}
    \end{equation}
With the aid of the above result, we can write the solution to the wave equation in the form of a Frobenius series, akin to Leaver's method~\cite{1985:Leaver}, to search for the quasinormal frequencies. This solution takes the form
\begin{equation}\label{EQ:appropriatepsiSdS}
    \begin{aligned}
		\psi(r)=\left(\frac{1}{r_{c}}-\frac{1}{r}\right)^{\rho _{c}}
		\left(\frac{1}{r}-\frac{1}{r_{h}}\right)^{\rho _h}
		\left(\frac{1}{r}+\frac{1}{r_{c}+r_{h}}\right)^{\rho_{h}-\rho_{c}}\\
    \times \sum_{n=0}^{\infty}a_{n}\left(\frac{\frac{1}{r}-\frac{1}{r_{h}}}{\frac{1}{r_{c}}-\frac{1}{r_{h}}}\right)^{n}.
    \end{aligned}
\end{equation}    
Substituting Eq.~(\ref{EQ:appropriatepsiSdS}) into Eq.~(\ref{EQ:diffpsiR}), we find a five-term recurrence relation for the $a_{n}$ coefficients
\begin{equation}\label{eq:recurrence}
 \begin{aligned}
	&	\alpha_{0}a_{1}+\beta_{0}a_{0}=0,\\
	&	\alpha_{1}a_{2}+\beta_{1}a_{1}+\gamma_{1}a_{0}=0,	\\	
	&	\alpha_{2}a_{3}+\beta_{2}a_{2}+\gamma_{2}a_{1}+\delta_{2}a_{0}=0,\\
    &   \alpha_{n}a_{n+1}+\beta_{n}a_{n}+\gamma_{n}a_{n-1}+\delta_{n}a_{n-2}+\sigma_{n}a_{n-3}=0, ~~ n\geq 3,
	\end{aligned}
\end{equation}
where the forms of $\alpha_{n}$, $\beta_{n}$, $\gamma_{n}$, $\delta_{n}$ and $\sigma_{n}$ are presented in Appendix \ref{AP:recurrencerelations}. Note that the expansion in Eq.~\eqref{EQ:appropriatepsiSdS} satisfies the boundary conditions at both horizons, provided that the recurrence relation~\eqref{eq:recurrence} is satisfied. 

To solve the recurrence relations, we can use a two-step Gaussian elimination procedure to reach a three-term recurrence relation. Details on this procedure can be found in Appendix \ref{AP:recurrencerelations}. Once a three-term recurrence relation is reached, one can proceed with the standard Leaver's method to obtain the QNF as the roots of the algebraic equation \cite{1985:Leaver,1990:Leaver}:
        \begin{equation}\label{EQ:CF3term}
           0= \dfrac{\beta''_{0}}{\alpha''_{0}}-\dfrac{\gamma''_{1}}{\beta''_{1}-}\dfrac{\alpha''_{1}\gamma''_{2}}{\beta''_{2}-}\dfrac{\alpha''_{2}\gamma''_{3}}{\beta''_{3}-}...
        \end{equation}
where the definitions of $\alpha''_n$, $\beta''_n$ and $\gamma''_n$ can be found in Appendix \ref{AP:recurrencerelations}. The roots of this equation are the quasinormal frequencies and they ensure the convergence of the series Eq.~(\ref{EQ:appropriatepsiSdS}).

\subsubsection{Schwarzschild Spacetime}
     In the Schwarzschild spacetime we follow similar steps to obtain both QNM and QBS \cite{2004:konoplyaZidenko,2007:SamDolan}. Since both spacetimes present spherical symmetry, the equations in Sec. \ref{sec:Introduction} have the same forms, with differences arising only at the level of the lapse function $f(r)$, since $\Lambda\to 0$ in Eq.~(\ref{EQ:lapsefunction}). The boundary condition at infinity for the case of a massive field in Schwarzschild spacetime needs to take into consideration the subdominant term at infinity due to the presence of mass. The appropriate Frobenius series is thus
     \begin{equation}\label{EQ:appropriatePSISchw}
         \begin{aligned}
         \psi(r)=\left(\frac{r}{r_{h}}-1\right)^{-\,i\,\omega\,r_{h}}e^{F(\omega)\,r}\left(\frac{r}{r_{h}}\right)^{\left(F(\omega)+i\,\omega-\frac{\mu^{2}}{2\,F(\omega)}\right)\,r_{h}} \\
         \times \sum_{n=0}^{\infty}a_{n}\left(\frac{r-r_{h}}{r}\right)^{n}
            \end{aligned}
     \end{equation}
 where $F(\omega)=\pm \sqrt{\mu^{2}-\omega^{2}}$. As mentioned after Eq.~\eqref{EQ:BoundaryPsiSdS} the choice of signal for ${\rm Re}[F(\omega)]$ determines the mode behavior at infinity. Indeed, if ${\rm Re}[(F(\omega)]>0$ it corresponds to quasinormal modes, whereas if ${\rm Re}[(F(\omega)]<0$ it corresponds to quasibound states. Substituting Eq.~(\ref{EQ:appropriatePSISchw}) into Eq.~(\ref{EQ:diffpsiR}) yields a three-term recurrence relation for the $a_{n}$ coefficients
 \begin{equation}\label{eq:recurrence2}
 \begin{aligned}
	&	\alpha_{0}a_{1}+\beta_{0}a_{0}=0,\\
	&	\alpha_{n}a_{n+1}+\beta_{n}a_{n}+\gamma_{n}a_{n-1}=0,~~ n\geq 1,
	\end{aligned}
\end{equation}
where $\alpha_n$, $\beta_n$ and $\gamma_n$ are given in Appendix \ref{AP:recurrencerelations}. Proceeding with the standard Leaver's method leads to an equation of the same form of Eq.~(\ref{EQ:CF3term}), without the primes, and solving numerically we find the QNF for ${\rm Re}[F(\omega)]>0$ or the or QBS for ${\rm Re}[F(\omega)]<0$, respectively.

At this point, it is interesting to highlight two properties of the equations until now. If we consider Eqs.~(\ref{EQ:BoundaryPsiSdS2})--(\ref{EQ:appropriatepsiSdS}), and take the limit $r_{c}\to\infty$, we recover the correspondent equations for the \textit{massless} scalar field in Schwarzschild spacetime, although there is no condition imposed on the mass of the field. This is a direct consequence of the potential behavior close to the cosmological horizon, as already mentioned after the Eq.~(\ref{EQ:BoundaryPsiSdS}). Further, since there are no assumptions on the mass of the field, the coefficients from the recurrence relation do not recover all the correspondent coefficients obtained from the asymptotic flat case.

\subsection{Direct integration method}

Another way to impose that the solutions satisfy the boundary conditions in Eq.~\eqref{EQ:BoundaryPsiSdS2} at the boundaries is to numerically integrate the equations under these conditions. This method is usually known as the direct integration (DI) method~\cite{chandradet,Macedo:2016wgh}, but it consists essentially of a shooting method for a two-point boundary value problem. We outline the procedure in what follows.

We start by numerically integrating the radial equation outward, starting from the event horizon, with the following boundary condition
\begin{equation}\label{eq:cond1}
    \psi(r\approx r_h)=\sum_{n=0}^{N}b_n (r-r_h)^n e^{-i\omega r_*}, 
\end{equation}
where the coefficients $b_n$ are found by expanding the equations near the event horizon and solving iteratively order by order in $(r-r_h)$. The minimum order of the expansion at the event horizon is chosen such that the computation of the mode frequency is numerically stable, which in our case corresponds typically to $N=5$. Therefore, we construct a numerically integrated solution, say $\psi_-$, from the event horizon, which satisfies the boundary condition at the event horizon but not necessarily at the cosmological horizon. 

The procedure outlined above can then be repeated for an inward integration starting from the cosmological horizon, under the boundary condition
\begin{equation}\label{eq:cond2}
    \psi(r\approx r_c)=\sum_{n=0}^{N}c_n (r-r_c)^n e^{i\omega r_*},
\end{equation}
where the $c_n$ coefficients are found by expanding the equations at the cosmological horizon and solving iteratively. The order of the expansions is typically set at $N=5$ as well, and checking convergence with other results in the literature. With this, we construct a second numerical solution, say $\psi_+$, that satisfies the boundary condition at the cosmological horizon.

Quasinormal mode solutions are obtained by requiring that the proper boundary conditions, namely Eqs.~\eqref{eq:cond1} and \eqref{eq:cond2}, are satisfied simultaneously. This is achieved by searching for the frequency for which the Wronskian $W$ of $\psi_-$ and $\psi_+$ vanishes, which implies that the solutions $\psi_-$ and $\psi_+$ are linearly dependent. The solutions can thus be extracted at an intermediate matching point $r=r_m$, at which one can find the roots $\omega$ of
\begin{equation}
    W(\psi_-,\psi_+)=\psi_-(r_m)\psi_+'(r_m)-\psi_-'(r_m)\psi_+(r_m)=0.
\end{equation}
We can verify if the mode is stable numerically by relaxing the number of coefficients considered in the expansions, as well as varying the matching point and values considered for the numerical horizons.

\subsection{Time evolution of initial data}

To study the time evolution of the scalar field and the influence of the mass in the time domain profile, we rewrite Eq.~(\ref{EQ:diffpsi}) in terms of the null coordinates, $du=dt-dr_{*}$ and $dv=dt+dr_{*}$, which leads to the following partial differential equation
\begin{equation}\label{EQ:diffPSINULLcoordinates}
   4 \frac{\partial^{2}\psi(u,v)}{\partial u \partial v}+V(r)\psi(u,v)=0.
\end{equation}
The integration is to be held in a $u$-$v$ grid, where we use the Gundlach-Price-Pulling discretization procedure \cite{1994:Gundlach}
\begin{equation}
    \psi(N)=\psi(W)+\psi(E)-\psi(S)-h^{2}V(S)\frac{\psi(W)+\psi(E)}{8}+\mathcal{O}(h^{4}),
\end{equation}
where $h$ is the length of the edge of the grid, and the letters $N$, $W$, $E$, and $S$ represent different points on the grid, namely $N=(u+h,v+h)$, $W=(u+h,v)$, $E=(u,v+h)$, and $S=(u,v)$ \cite{2011:KonoplyaZhidenko}. The initial data are assumed to be a Gaussian wave packet in the $v=v_{0}$ null surface, and constant in the $u=u_{0}$, of the form
\begin{equation}
    \psi(0,v)=A_0\, {\rm exp}[-(v-v_c)^2/\sigma],
\end{equation}
with $(A_0,v_c,\sigma)$ being constants. For this work, we have considered $(A_0,v_c,\sigma)=(1,10,2)$. Modifying the initial condition does not change the time domain profile of the perturbation in the ringdown phase nor in the late time. To obtain the profile, we extract the field at $r_{*}=10 r_{h}$ (for other values the behavior remained consistent with our results)~\footnote{We observe that the tortoise coordinate is defined up to an arbitrary constant. We define the tortoise coordinate such that $r_*(r=1.27 r_{h})=0$ is close to the point at which the potential is maximum for the cases we consider.}.
We use $h=5\times10^{-1}$ for the grid, verifying that smaller values do not change the results considerably, and also testing convergence in some particular points. Finally, we use a Prony method to find the dominant QNF that appears in the ringdown phase and in the late time (see, for instance, Ref.~\cite{2011:KonoplyaZhidenko}),
fitting the signal with the series expansion
\begin{equation}
    \psi=\sum_{i=0}^N C_{i}~e^{-i\omega_i t},
\end{equation}
and finding the (complex-valuated) coefficients $(C_i,\omega_i)$.

\section{Results}\label{sec:results}

\begin{figure}[h]
    \includegraphics[width=1\linewidth]{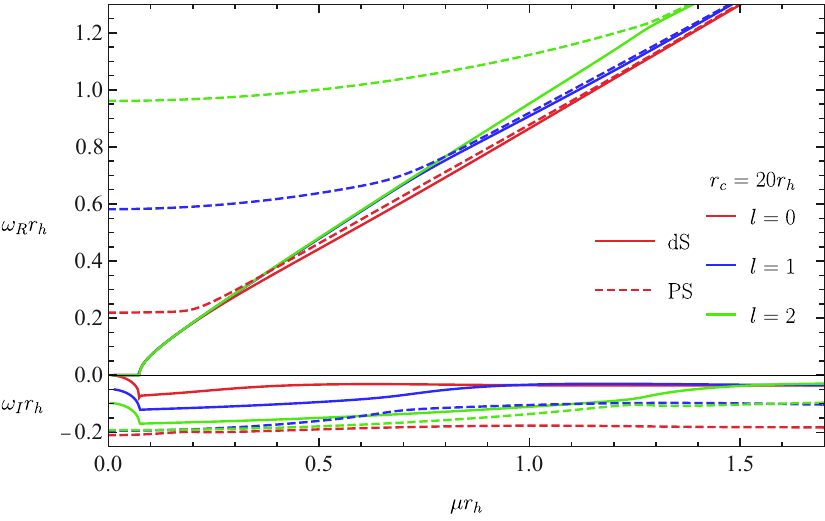}
    \caption{Real and imaginary part of the fundamental PS and dS quasinormal modes of massive scalar fields in SdS BHs as a function of the mass of the field. The PS modes are represented by dashed lines and the dS modes by solid lines. We show the cases for $l=\{0,1,2\}$ and $r_{c}=20\,r_h$, but the results qualitatively hold for other values of $r_c$.}
\label{FIG: QNMF_l0l1_rc20_VARmu}
\end{figure}

In SdS BHs there are two families of modes. The first family is the PS modes, whose frequencies are denoted by $\omega^{PS}$, and it asymptotically approaches the usual Schwarzschild QNMs as $r_{c}\to \infty$. These are related to the unstable null circular geodesics and can also be determined by the WKB method \cite{2017:ToshmatovStuchlik}. On the other hand, the dS modes, $\omega^{dS}$, result from the de Sitter behavior of the system and can be purely imaginary, depending on the mass of the field. Using the methods described in Sec.~\ref{sec:qnms}, we compute the QNFs and compare our results with the ones presented in the literature for massless scalar fields~\cite{2020:Aragon}. Our results are presented in detail in Appendix~\ref{app:massless}. The massive case shares some features with the massless case, but there are some crucial differences introduced in the massive case, as expected given the modifications induced in the effective potential (see Fig.~\ref{FIG:PotentialVARlambdaMu}).

\begin{figure}[t!]
    \includegraphics[width=1\linewidth]{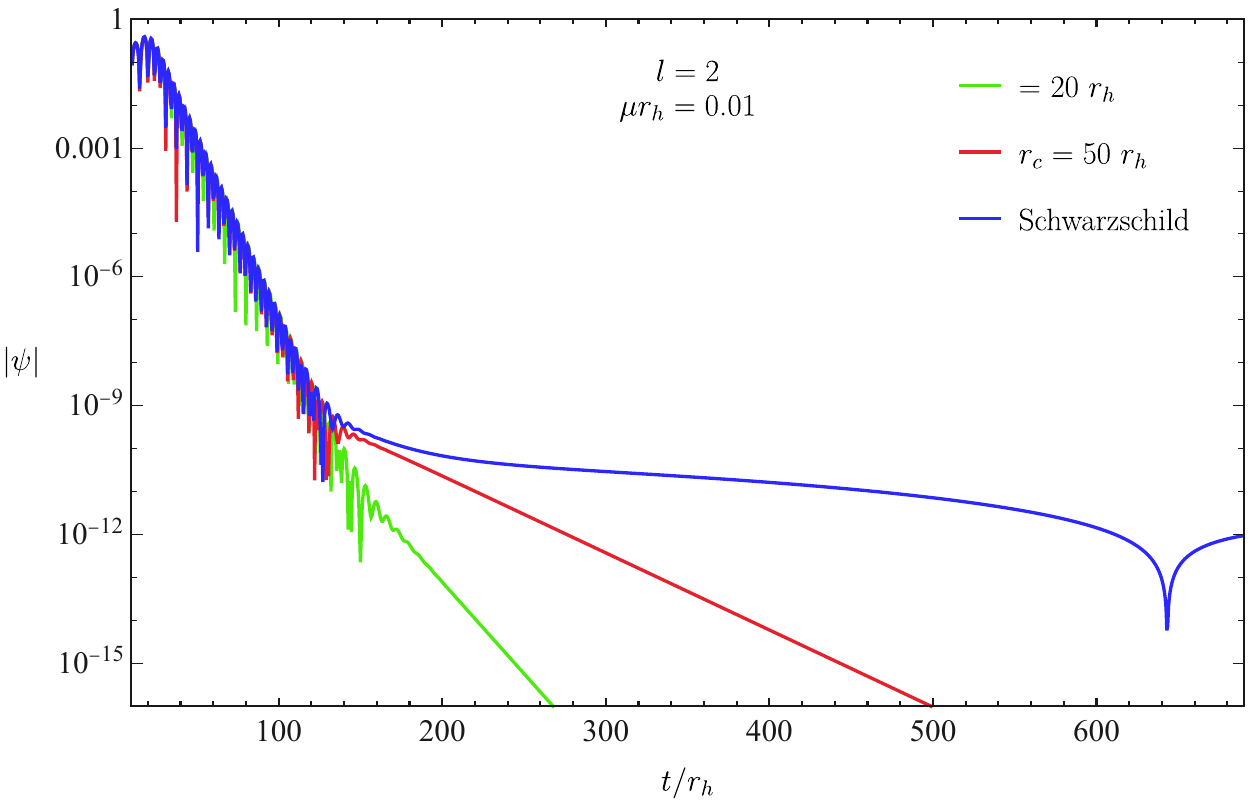}
    \caption{Time evolution of scalar wave packets  for $\mu r_{h}=0.01$ and $l=2$, and different values of the cosmological horizon ($r_{c}$), including the Schwarzschild case ($\Lambda=0$). A nonzero cosmological constant removes the quasibound states part of the spectrum and, in these cases, the dS modes are purely imaginary.
    }
    \label{FIG:TEV_l2_varRCmu001_varMUrc20}
\end{figure}

In Fig.~\ref{FIG: QNMF_l0l1_rc20_VARmu}, we plot the QNFs as functions of the mass of the field for a constant value of the cosmological constant. The behavior of the modes can be qualitatively divided into three parts. For $\mu<\mu_{c}$, for some critical value $\mu_c$ the dS mode frequencies are purely imaginary and, consequently, after the ringdown we have an exponential ``tail'' corresponding to this mode. The PS modes vary only slightly in this region. The value of $\mu_{c}$ depends on the cosmological constant, and for the de Sitter case (with $M=0$) they correspond to
\begin{equation}\label{EQ:mulimite}
    \mu_{c}^{2}=\dfrac{3}{4} \Lambda.
\end{equation}
Although such an explicit form is not attainable for the SdS case, the result for the de Sitter spacetime presented in Ref.~\cite{2006:Lopez-Ortega} and \cite{2020:Aragon} provides a good estimate for our purposes. In this region, the imaginary part of the dS modes increases in absolute value, implying that as $\mu\rightarrow \mu_{c}$ the decay timescale decreases. 

\begin{figure*}[th]
    \includegraphics[width=.475\linewidth]{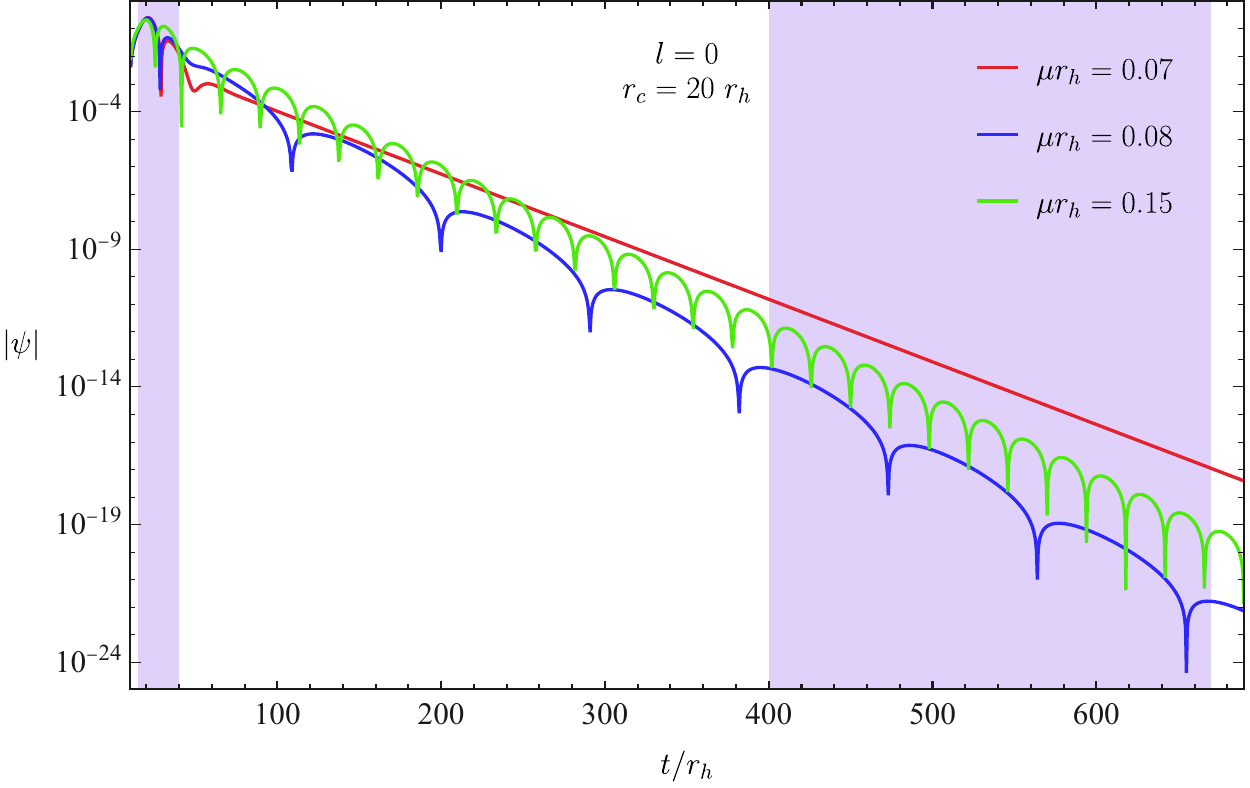}\qquad
    \includegraphics[width=.475\linewidth]{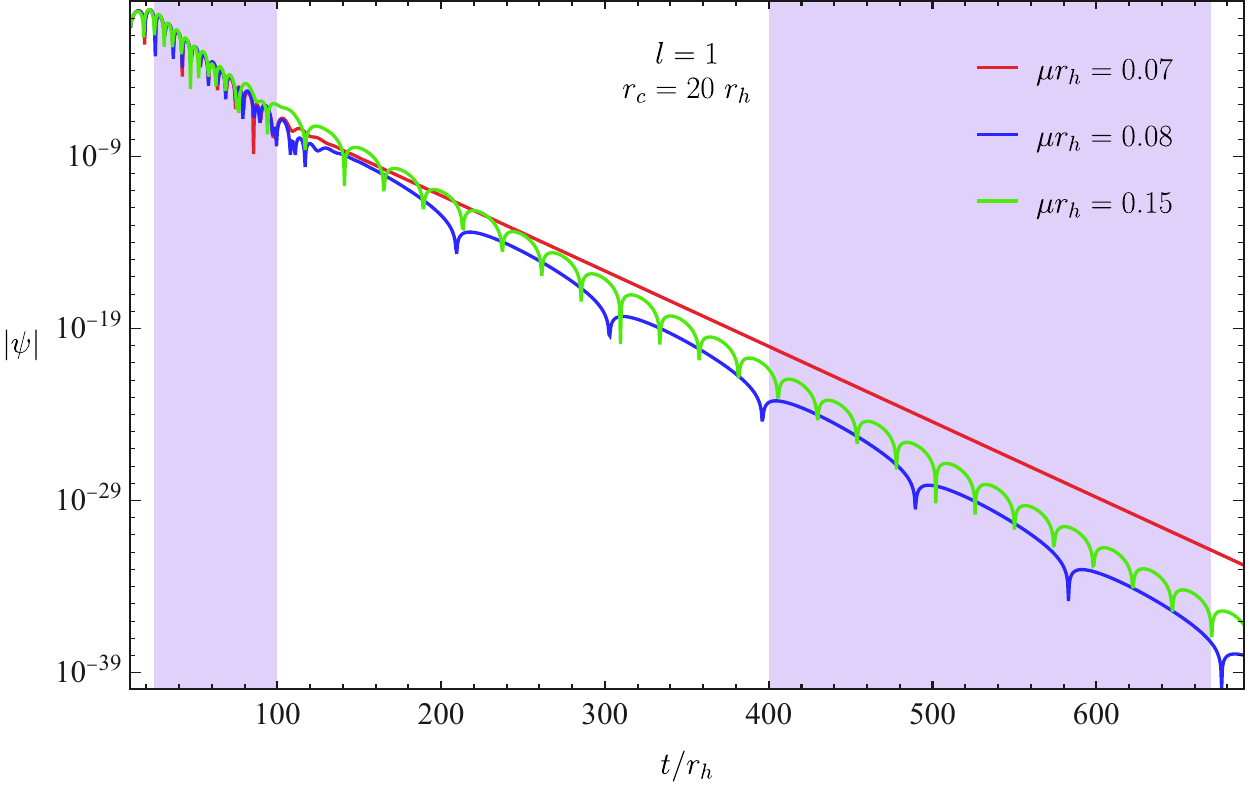}
    \caption{Time domain profile of the scalar field for $l=0$ (left panel) and $l=1$ (right panel), for different scalar field masses. The field's masses are chosen such that each value highlights the different behaviors of the tail as we cross the limit for the purely imaginary modes. The shaded region indicates the interval in which the frequencies are extracted using the Prony method.}
    \label{FIG:TE_l0l1_VARMu}
\end{figure*}
\begin{table*}[t!]
    \caption{\label{tab:QNMfreqTE_varmu} Fundamental QNF for the configurations of Fig. \ref{FIG:TE_l0l1_VARMu}, using the Leaver and Prony method.}
    \begin{ruledtabular}
    \begin{tabular}{|c|c|c|c|c|c|}

			$l$ &$\mu r_{h}$ & $\omega_{Leaver}^{dS}r_{h}$   & $\omega_{Prony}^{dS}r_{h}$ & $\omega_{Leaver}^{PS} r_{h}$ & $\omega_{Prony}^{PS} r_{h}$\\ 
			\hline                              
		&	$0.07$ & $-0.0523507\,i$  & $-0.052711 \,i$ & $0.219859 - 0.208169\,i$ & $0.24547 - 0.22047 \,i$\\
	0	&	$0.08$ & $0.034485 - 0.0715712\,i$  & $0.034621 - 0.071687\,i$ & $0.220026 - 0.20755\,i$ & $0.25084 - 0.21524\,i$\\
		&	$0.15$ & $0.130808 - 0.0643422\,i$  & $0.13113 - 0.06447\,i$ & $0.222553 - 0.201355\,i$ & $0.30933 - 0.19301\,i$\\
  \hline
		& $0.07$& $-0.100924\,i$ &   $-0.10115\,i$ & $0.582771 - 0.194107\,i$ & $0.59178 - 0.19401\,i$\\
        1    & $0.08$& $0.0336226 - 0.120839\,i$ &  $0.033702 - 0.121088\,i$ & $0.583106 - 0.193909\,i$ &  $0.59532 - 0.19470\,i$  \\
            & $0.15$& $0.130563 - 0.117177 \,i$ &  $0.13071 - 0.11733\,i$ & $0.586701 - 0.191773\,i$ &  $0.59104 - 0.19245\,i$ \\

		\end{tabular}
         \end{ruledtabular}
	\end{table*}

For $\omega_{R}^{PS} 	\gtrsim \mu > \mu_{c}$, as we show in what follows, the time domain profile can be separated between a clear "photon sphere" ringdown and a further decay dominated by the dS modes, with oscillations that resemble that of a massive field in Schwarzshild spacetime. In this regime, the dS modes have both real and imaginary parts. At intermediate times, the signal presents some interference due to the influence of both PS and dS modes. The decay rate increases with $\mu$, reaching a maximum at some value near $\mu_c$ that depends on $r_c$, and then starts to decrease. For example, for $r_{c}=20 r_{h}$ the decay rate decreases for masses $\mu>0.08\, r_{h}^{-1}$.

The third region occurs for $\mu>\omega_{R}^{PS}$
. In this region, the PS modes have a significant change. As we can see in Fig.~\ref{FIG: QNMF_l0l1_rc20_VARmu}, the real part of both PS and dS modes is bounded by the mass of the field, i.e. $\omega_{R}^{PS,dS} <\mu$, and they increase approximately linearly with $\mu$, converging to similar values in the large $\mu$ limit. The imaginary part of the dS modes approaches a constant value regardless of the multipole $l$. For instance, for $r_c=20r_h$, $\omega_I^{dS}r_{h}\sim-0.03633$ for large $\mu$.
As the real part of the PS and dS modes is similar in the large $\mu$ limit, the main distinction between them comes from the imaginary part. The time domain profile of the field ceases to present an interference profile, and the dS modes dominate the signal as we can see through the Prony method. 

Note that the transitions between the above regions occur for considerably large values of the mass of the field. For instance, we can see in Fig.~\ref{FIG: QNMF_l0l1_rc20_VARmu} that for $l=0$ and $r_{c}=20r_{h}$ the first transition happens at $\mu r_{h}\sim 0.073$, and the second one at $\mu r_{h}\sim\omega_{R} r_{h}\sim 0.22$.

To illustrate the features explained above in the time evolution of the signals, in Fig.~\ref{FIG:TEV_l2_varRCmu001_varMUrc20} we present the time domain profile of the massive scalar field with the angular number $l=2$. In particular, we consider a massive scalar field with $\mu r_{h}=0.01$ in the SdS with the cosmological horizon located at $r_{c}=20 r_{h}$ ($\mu_c r_h\approx 0.073$) and $50 r_{h}$ ($\mu_c r_h\approx 0.03$) and for the asymptotic flatness case ($\Lambda=0$). Notice that in the cases with cosmological constant we have $\mu<\mu_c$ and, therefore, the dS modes are purely imaginary. We see that there is a late-time oscillation introduced by the mass of the field in the asymptotically flat case, as usually is the case for massive fields in Schwarzschild. For the SdS cases, as $\mu<\mu_c$, the exponential late-time decay is dictated by the purely imaginary dS mode. We highlight that the ringdown phase of the signals presented in Fig.~\ref{FIG:TEV_l2_varRCmu001_varMUrc20} is essentially the same, which is consistent with the fact that the PS mode frequencies are approximately the same for $\mu<\mu_c$.

In Fig. \ref{FIG:TE_l0l1_VARMu} we show the time domain profile of the perturbation for fixed cosmological horizon $r_{c}=20 r_{h}$ ($\mu_c r_h\approx0.073$). In the left panel (right panel) we plot the $l=0$ ($l=1$) mode, for different values of the scalar field mass $\mu r_{h}=\{0.07,0.08,0.15\}$. In both cases, we see the dS modes acquiring an oscillatory pattern for $\mu>\mu_c$, indicating that the modes acquire a real part. We also see an increasing decaying rate, in agreement with the behavior previously discussed. The corresponding QNM frequencies for the configurations in Fig. \ref{FIG:TE_l0l1_VARMu} are presented in Table \ref{tab:QNMfreqTE_varmu}. We show the numbers found through both the Leaver's and Prony methods, with the direct integration method computations in a close agreement with those of the Leaver's method. The inaccuracy of the Prony method for $l=0$ is caused by the small number of oscillations present in the ringdown phase.

\begin{figure*}[t!]
    \includegraphics[width=.48\linewidth]{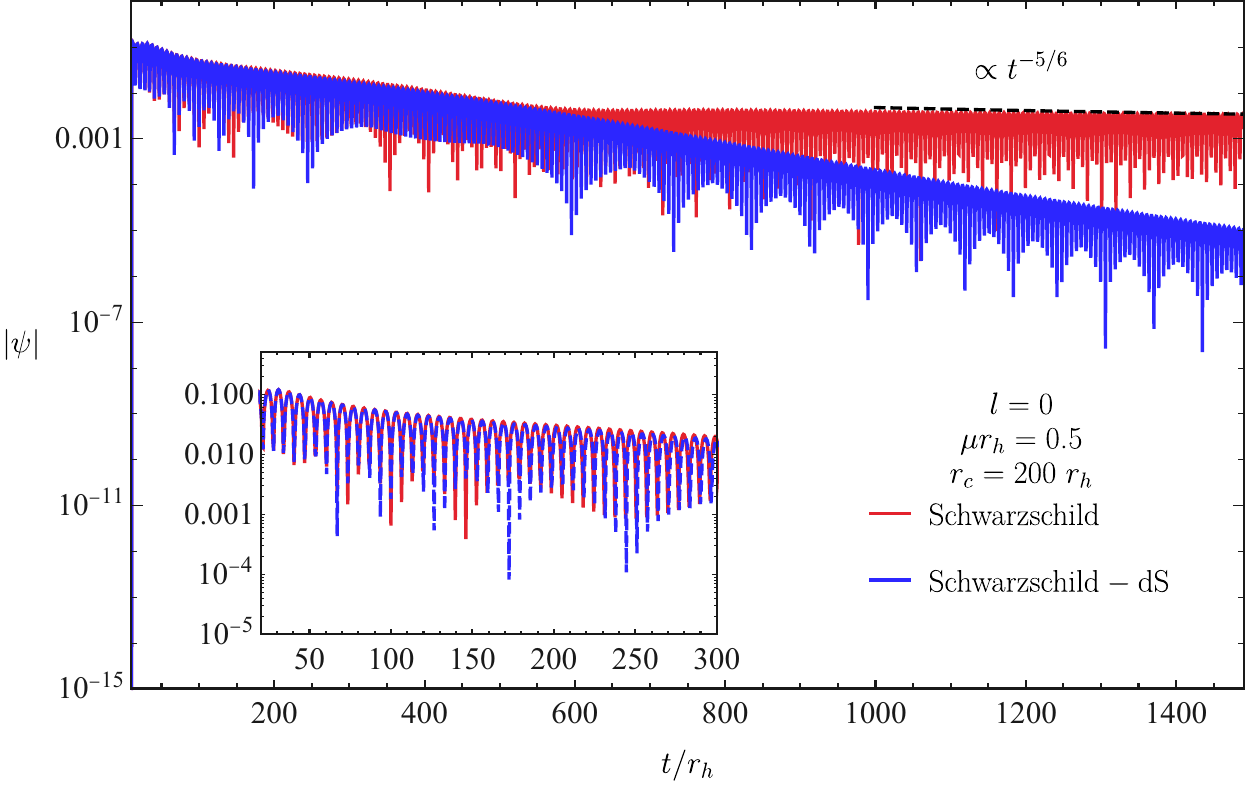}
    \includegraphics[width=.48\linewidth]{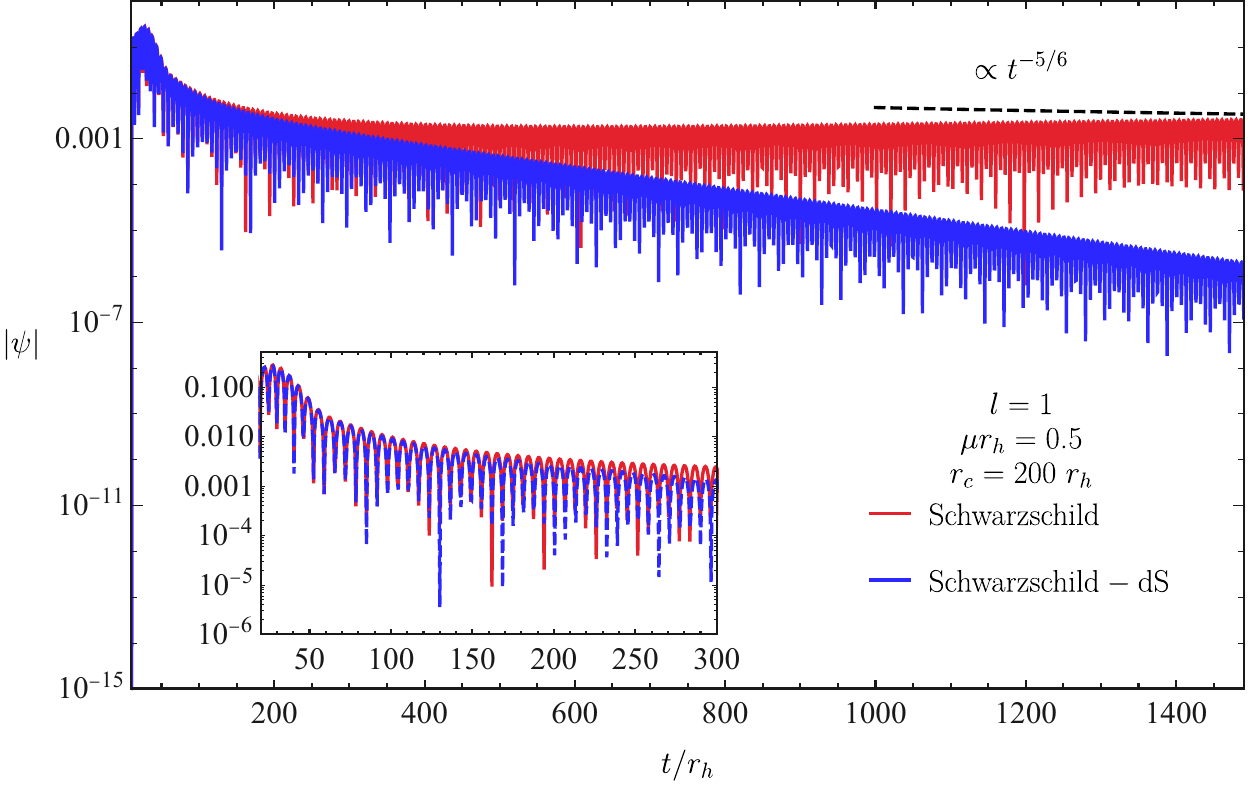}
    \caption{Time evolution of a massive scalar field with a mass $\mu r_{h}=0.5$, in Schwarzschild and Schwarzscild-de Sitter with $r_{c}=200 r_{h}$, for $l=0$ (left panel), and $l=1$ (right panel). We see that the time evolution is comparable in early stages, but the decay in SdS is quicker, while in the Schwarzshild case it behaves as $t^{-5/6}$ at late times.}
    \label{FIG:SvsSdS_TE_l0l1}
\end{figure*}

It is relevant to investigate whether there is a limit at which the signal of SdS BHs coincides with their asymptotically flat counterparts. To access that, in Fig. \ref{FIG:SvsSdS_TE_l0l1} we compare the time evolution of massive scalar fields in Schwarzschild with the SdS for a choice of cosmological horizon radius large in comparison with the BH length scale ($r_{c}=200 r_{h}$). We also consider a large value of the scalar field mass, $\mu r_{h}=0.5$, to analyze whether quasibound features appear in SdS BHs. For completeness, we show the cases $l=0$ (left panel) and $l=1$ (right panel). Initially, the profiles are almost identical, even for moderate time intervals (see also the inset in Fig.~\ref{FIG:SvsSdS_TE_l0l1}). However, at late times the universal behavior of massive fields in Schwarzschild BHs ($t^{-5/6}$) changes when the cosmological constant is present. For SdS BHs the late time is described by the dS modes. 

\begin{figure}[t!]
\includegraphics[width=1\linewidth]{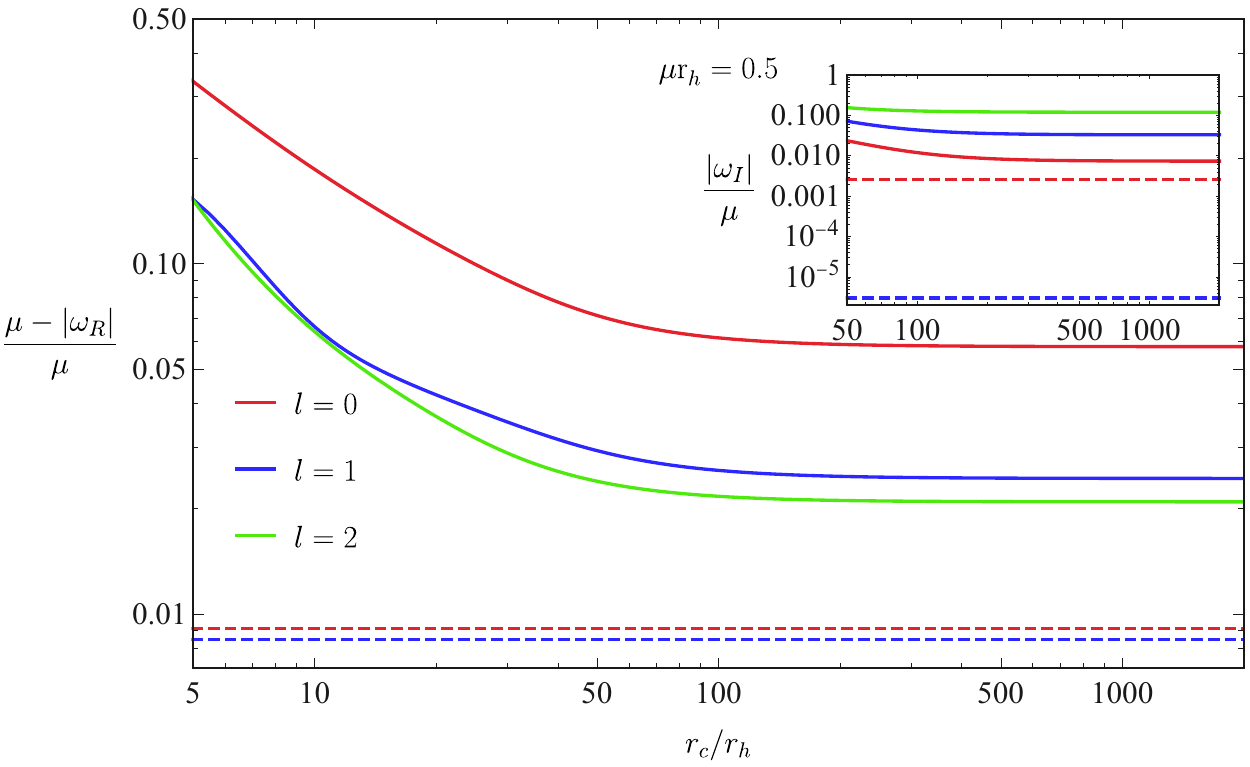}
   \caption{Quasinormal frequencies of dS modes (solid lines) and the quasibound frequencies in Schwarzschild spacetime (dashed lines) for $\mu r_{h}=0.5$, different values of $l=0$, $1$ and $2$, and increasing the cosmological horizon coordinate. }
    \label{FIG:dSM_varRC}
\end{figure}

 \begin{table*}[t!]
    \caption{\label{tab:results RCINFdSmodes} Quasinormal frequencies of de Sitter modes for $r_c=2000r_h$, in the limit $r_{c} \to \infty$, and the quasibound states in Schwarzschild spacetime ($\omega^{QBS}$) for different values of angular number $l$ and for $\mu r_{H}= 0.5$.}
    \begin{ruledtabular}
    \begin{tabular}{|c|c|c|c|}

			$l$ & $\omega_{Leaver}^{dS}/ \mu$ ($r_{c}=2000 r_{h}$) & $\omega_{Leaver}^{dS}/\mu$ ($r_{c}\rightarrow \infty$) & $\omega^{QBS}/\mu$\\ 
			\hline                              
			$0$ & $0.960624 - 0.0154134\,i$ &$0.96064 - 0.0153784\,i$ & $0.990915 - 0.0025461\,i$ \\
			$1$ & $0.984614 - 0.00390827\,i$ &$0.984633 - 0.00382118\, i$ & $0.991565 - 2.97451\times 10^{-6}\,i$\\
			$2$ & $0.987621 - 0.0174992\,i$&$0.98763 - 0.0173872\, i$  & $0.99644 - 9.01895\times 10^{-12}\,i$\\
		\end{tabular}
         \end{ruledtabular}
	\end{table*}

It is important to mention that for small values of the cosmological constant (or $r_{c}\gg r_h$) the contributions of PS and dS modes are more clearly distinguishable. As $r_{c}\rightarrow r_{h}$, the critical mass is $\mu_cr_h\approx0.87$, and therefore the range of mass of the field in which the dS modes have real part is relatively large. For such larger masses, the time domain profile does not present clear distinguishable features from the PS and dS modes.

The frequencies of the dS modes have a direct relation with the tail present in the time evolution and also with the mass of the field. When $\mu>\mu_{c}$, we have $\omega_{R}^{dS}\neq 0$, and oscillations appear in the tail. Furthermore, the condition $\omega_{R}^{dS}/\mu <1$ is always verified. These are properties similar to the case of quasibound states of a massive scalar field in the Schwarzschild spacetime \cite{2007:SamDolan,2004:BurkoANDKhanna}. From these similarities, we might interpret the dS modes as some sort of reminiscent of quasibound states from the asymptotic flat case, being more and more similar in the limit $r_c\to \infty$.  To investigate the validity of this statement, we track the dS modes as we approach the asymptotic flat case for a fixed value of the mass $\mu r_h=0.5$, and the result can be seen in Fig.~\ref{FIG:dSM_varRC}. These results indicate that there is always a gap between the quasibound and the dS modes, which is smaller for larger values of $l$. Particularly, we have calculated $\omega^{dS}$ through two different methods, namely by considering the dS modes for $r_{c}=2000r_{H}$ and by taking the limit $r\to\infty$ in the coefficients of the continued fraction in Appendix \ref{AP:recurrencerelations}. The results can be seen in Table~\ref{tab:results RCINFdSmodes} together with the fundamental quasibound states in Schwarzschild spacetime. We highlight that, although the spectrum of dS modes does not coincide with the spectrum of quasibound states in the $r_c\to\infty$ case, the time evolution shares similar profiles, especially at early times, with the dS modes appearing in the late-time evolution.

\section{Conclusion}\label{sec:conclusion}
    In the present paper, we have studied the evolution of a massive scalar field in the background of a Schwarzschild-de Sitter black hole. We determined the quasinormal frequencies using the Leaver, Chandrasekhar, and Prony methods. We found two branches of frequencies, the photons sphere modes, which are related to the circular null geodesics, and the de Sitter modes which seem to be related to the asymptotic potential behavior. 
    
    The photon sphere modes for large values of the cosmological horizon dominate the ringdown phase of the time evolution. Increasing the mass of the scalar field mainly increases its oscillation behavior and increases its decay time. For small values of the cosmological horizon, these modes dominate the whole time evolution of the perturbation, overshadowing any contribution from the de Sitter modes.

    The de Sitter modes increase in magnitude for small values of $r_{c}$, but their main influence is better observed for greater values of the cosmological horizon. For a massless scalar field, the dS modes are purely imaginary and can be extracted through the Prony method from the exponential tail. Increasing the mass of the scalar field, these modes can acquire a real part, which changes the tail, introducing some oscillations. As we increase the mass of the field, we initially see these modes decreasing the time decay. Still, once a threshold is crossed, the imaginary part of the dS modes decreases, giving rise to modes that are similar to the long-lived modes in the Schwarzschild asymptotic flat case.

    To verify whether the de Sitter modes are reminiscent of the quasibound states, we track them as we increase the cosmological horizon to recover the asymptotic flatness. We found that, for the range of parameters considered, the de Sitter modes share similar properties with the quasibound states, although their spectrum does not retrieve the quasibound state spectrum in the limit $r_{c}\to\infty$.

We also note here the similarities between massive fields in de Sitter black holes and the effective mass of the photon introduced by astrophysical plamas~\cite{Cardoso:2020nst,Wang:2022hra}. Since the plasma is localized, the effective mass of the field does not introduce a quasibound spectrum. Nonetheless, dynamical evolution shows a similar time-domain profile to the true massive case at early stages, akin to what we presented in this paper. It would be interesting to analyze common features in both scenarios.

Finally, we note that the natural extension of our results would be to consider spinning black holes. In the rotating case, the mass of the field is connected with superradiant instabilities~\cite{2020:BritoCardosoPani} and the existence of a cosmological horizon could quench such instabilities since the field is not confined anymore. Therefore, the existence of a cosmological horizon would potentially relax the bounds on the rotation of black holes as well as on the mass of ultralight bosons~\cite{Pani:2012vp,Brito:2013wya}. We leave this for future work.
    
\section{acknowledgments}
We thank Hector O. Silva for the discussions. C. F. B. M. and M. M. C. acknowledge Fundação
Amazônia de Amparo a Estudos e Pesquisas (FAPESPA),
Conselho Nacional de Desenvolvimento Científico e Tecnológico (CNPq) and Coordenação de Aperfeiçoamento de 
Pessoal de Nível Superior (CAPES) – Finance Code 001,
from Brazil, for partial financial support. J. L. R. acknowledges the European Regional Development Fund and the program Mobilitas Pluss for financial support through Projects No.~MOBJD647, and No.~2021/43/P/ST2/02141 cofunded by the Polish National Science Centre and the European Union Framework Programme for Research and Innovation
Horizon 2020 under the Marie Sklodowska-Curie Grant Agreement No. 94533, Fundação para a Ciência e Tecnologia through Project No. PTDC/FIS-AST/7002/2020, and Ministerio de Ciencia, Innovación y Universidades (Spain), through Grant No. PID2022-138607NB-I00.

\section{APPENDIX A: Recurrence Relations and Gaussian Elimination}\label{AP:recurrencerelations}
\subsubsection{Schwarzschild case}

In Sec. \ref{SUBSEC:COntFrac}, we found a three-term recurrence relation for the Schwarzchild case, see Eq.~\eqref{eq:recurrence2}. The coefficients $\alpha_{n}$, $\beta_{n}$ and $\gamma_{n}$ are given explicitly by
\begin{widetext}
\begin{equation}
		\alpha_{n}=(1+n)(1+n-2\,i\,r_{h}\omega),
	\end{equation}
	\begin{equation}
		\begin{aligned}
			\beta_{n}=\frac{(2 n+1) r_h \left(3 \mu ^2+4 i \omega  (F(\omega )+i \omega )\right)}{2 F(\omega )}+\left(\frac{i \omega ^3}{F(\omega )}-2 i \omega  F(\omega )-F(\omega )
   (F(\omega )+i \omega )+3 \omega ^2\right) r_h^2-l (l+1)-2 n (n+1)-1,
	\end{aligned}
	\end{equation}

	\begin{equation}
		\begin{aligned}
		\gamma_{n}=-\frac{n r_h \left(\mu ^2+2 i \omega  (F(\omega )+i \omega )\right)}{F(\omega )}+\frac{r_h^2 \left(4 i \mu ^2 \omega  (F(\omega )+2 i \omega )+8 \omega ^3 (\omega -i
   F(\omega ))+\mu ^4\right)}{4 F(\omega )^2}+n^2,
		\end{aligned}
	\end{equation}
where $F(\omega)=\pm\sqrt{\mu^{2}-\omega^{2}}$. The ratio between successive series coefficients $a_{n}$ leads to an infinite continued fraction
\begin{equation}
		0=\beta_{0}-\dfrac{\alpha_{0}\gamma_{1}}{\beta_{1}-\dfrac{\alpha_{1}\gamma_{2}}{\beta_{2}-\dfrac{\alpha_{2}\gamma_{3}}{\beta_{3}-...}}},
	\end{equation}
 which can be rewritten as
\begin{equation}\label{EQ:ContFracApendixSChw}
           0= \dfrac{\beta_{0}}{\alpha_{0}}-\dfrac{\gamma_{1}}{\beta_{1}-}\dfrac{\alpha_{1}\gamma_{2}}{\beta_{2}-}\dfrac{\alpha_{2}\gamma_{3}}{\beta_{3}-}...
    \end{equation}
    We solve this numerically to find the QNF or QBS as the roots of the Eq.~(\ref{EQ:ContFracApendixSChw}).

    \subsubsection{Schwarzschild-de Sitter case}
     In the SdS case, we found a five-term recurrence relation given by Eq.~(\ref{eq:recurrence}), the full form of the coefficients $\alpha_{n}$, $\beta_{n}$, $\gamma_{n}$, $\delta_{n}$, and $\sigma_{n}$ are
\begin{equation}
\alpha_{n}=\frac{2\,M(1+n)(r_{c}+2\,r_{h})}{r^{3}_{h}(r_{c}+r_{h})}(1+n+2\,\rho_{h}),
	\end{equation}
	\begin{equation}
		\begin{aligned}
			\beta_{n}=-\mu^{2}-\frac{l(l+1)}{r^{2}_{h}}+&\frac{2\,M}{r^{3}_{h}r_{c}(r_{c}+r_{h})}\left((r^{2}_{c}+r_{c}r_{h})(1+2\,n(1+2\,n)+4\rho_{h}(1+3\,n+2\,\rho_{h}))\right)+ \\
			-&\frac{2\,M}{r^{3}_{h}r_{c}(r_{c}+r_{h})}r^{2}_{h}\left(2+n(1+5\,n)+2\,\rho_{h}(1+6\,n+2\,\rho_{h})\right),
	\end{aligned}
	\end{equation}

	\begin{equation}
		\begin{aligned}
		\gamma_{n}=\frac{2\,l (l+1) \left(r_c-r_h\right)}{r_c r_h^2}+\frac{2\,M \left(r_c-r_h\right) \left(r_c
			r_h+r_c^2\right) \left(22\,n\,\rho_h +20\, \rho_h^2-10 \rho_h+6 (n-1) n+3\right)}{r_c^2 r_h^3 \left(r_c+r_h\right)}+ \\
		-\frac{4\, M \left(r_c-r_h\right)r_h^2 \left(2
		\rho _h+n-1\right) \left(2\,\rho _h+2\,n-1\right)}{r_c^2 r_h^3 \left(r_c+r_h\right)},
		\end{aligned}
	\end{equation}
 
	\begin{equation}
		\begin{aligned}
		\delta_{n}=-\frac{l (l+1) \left(r_c-r_h\right){}^2}{r_c^2 r_h^2}-\frac{2\, M \left(r_c-r_h\right){}^2 \left(r_c\, r_h+r_c^2\right) \left(2\,n \left(8\,\rho
			_h-5\right)+4\,\rho _h \left(4\,\rho _h-5\right)+4\, n^2+7\right)}{r_c^3\,r_h^3
			\left(r_c+r_h\right)}+ \\
	+	\frac{2\,M \left(r_c-r_h\right){}^2 \left(2\,\rho _h+n-2\right) \left(2\,\rho _h+n-1\right)}{r_c^3\,
			r_h \left(r_c+r_h\right)},
	\end{aligned}
	\end{equation}
	\begin{equation}
		\sigma_{n}=\frac{2\,M(r_{c}-r_{h})^{3}}{r^{3}_{c}r^{3}_{h}}(2\,\rho_{h}+n-2)^{2}.
	\end{equation}
 \end{widetext}
To calculate the quasinormal modes we need to perform a two-step Gaussian elimination to reach a three-term recurrence relation \cite{1990:Leaver,2011:KonoplyaZhidenko}. The first step is to define new coefficients given by
\begin{equation}
    \alpha'_{n}\equiv \alpha_{n}, ~~ \beta'_{n}\equiv \beta_{n}, ~~ \gamma'_{n}\equiv\gamma_{n}, ~~ \delta'_{n}\equiv \delta_{n}, ~~ \text{for } n=0,\,1,
\end{equation}
and 
\begin{equation}
\begin{aligned}
    &\alpha'_{n}\equiv \alpha_{n}, ~~ \beta'_{n}\equiv \beta_{n}-\frac{\alpha'_{n-1}}{\delta'_{n-1}}\sigma_{n},  \\
    &\gamma'_{n}\equiv\gamma_{n}-\frac{\beta'_{n-1}}{\delta'_{n-1}}\sigma_{n}, ~~ \delta'_{n}\equiv \delta_{n}-\frac{\gamma'_{n-1}}{\delta'_{n-1}}\sigma_{n}, ~~ \sigma'_{n}\equiv 0 ~~ \text{for } n \geq 2.
\end{aligned}
\end{equation}
After this first step, we obtain a similar recurrence relation, but now with four terms, $\alpha'_{n}$, $\beta'_{n}$, $\gamma'_{n}$, and $\delta'_{n}$. We perform a second Gaussian elimination, defining the coefficients
\begin{equation}
    \alpha''_{n}\equiv \alpha'_{n},~~ \beta''_{n}\equiv \beta'_{n}, ~~ \gamma''_{n}\equiv\gamma'_{n}, ~~ \text{for } n=0,\,1,
\end{equation}
and
\begin{equation}
\begin{aligned}
   & \alpha''_{n}\equiv \alpha'_{n}, ~~ \beta''_{n}\equiv \beta'_{n}-\frac{\alpha''_{n-1}}{\gamma''_{n-1}}\delta'_{n}, \\
   & \gamma''_{n}\equiv\gamma'_{n}-\frac{\beta''_{n-1}}{\gamma''_{n-1}}\delta'_{n}, ~~ \delta''_{n}\equiv 0, ~~ \text{for } n\geq 2.
\end{aligned}
\end{equation}
These new coefficients obey a three-term recurrence relation
\begin{equation}\label{eq:recurrenceApendix}
 \begin{aligned}
	&	\alpha''_{0}a_{1}+\beta''_{0}a_{0}=0,\\
	&	\alpha''_{n}a_{n+1}+\beta''_{n}a_{n}+\gamma''_{n}a_{n-1}=0, ~~ n\geq 1.
	\end{aligned}
\end{equation}
Now with a three-term recurrence relation we can proceed in the same way as the Schwarzschild case. Equation (\ref{EQ:appropriatepsiSdS}) is a convergent series for the characteristic frequency $\omega$ that solves the equation Eq.~(\ref{EQ:ContFracApendixSChw}) for this case.
        
\section{APPENDIX B: Massless scalar modes}~\label{app:massless}
For consistency check, we compare the quasinormal frequencies, both PS (Table~\ref{tab:Comparingresults}) and dS (Table~\ref{tab:Comparingresults2}) modes, obtained through the Leaver's method and the DI with previous results in the literature for lower overtones \cite{2020:Aragon,2004:konoplyaZidenko}. The Leaver's method is also consistent with the high overtones presented in \cite{2017:Jansen}, but, since this work aims at analyzing the dominant modes and possible long-lived modes, we do not include a table for comparison.

 \begin{table*}[t!]
    \caption{\label{tab:Comparingresults} Comparing the quasinormal frequencies for the photon sphere modes ($n_{PS}=0$) for massless scalar field with $l=1$, from Ref. \cite{2020:Aragon} ($\omega_{PS}$), with our results obtained through Leaver ($\omega_{Leaver}$) and direct integration ($\omega_{DI}$).}
    \begin{ruledtabular}
    \begin{tabular}{|c|c|c|c|}

			$\Lambda M^{2}$ & $\omega_{PS}M$ \cite{2020:Aragon,2004:konoplyaZidenko} & $\omega_{Leaver}M$ &$\omega_{DI}M$ \\ 
			\hline                               
			$0.02$ & $0.2603-0.0911i$  & $0.26028785 - 0.091002536\,i$ & $0.260288 - 0.0910025\,i$ \\
			$0.04$ & $0.2247-0.0821i$  & $0.22468492 - 0.082051288\,i$ & $0.224685 - 0.0820513\,i$ \\
			$0.06$ & $0.1854-0.0701 i$  & $0.18536931 - 0.070060118\,i$ & $0.185369 - 0.0700601\,i$\\
			$0.08$ & $0.1404-0.0542i$  & $0.14040900 - 0.054023195\,i$ & $0.140409 - 0.0540232\,i$\\
			$0.09$ & $0.11392-0.04397i$  & $0.11399675 - 0.043882695\,i$ & $0.113997 - 0.0438827\,i$\\
			$0.1$ & $0.08156-0.03121i$  & $0.081589747 - 0.031233161\,i$ & $0.0815901 - 0.0312332\,i$ \\
			$0.11$& $0.02549-0.00965i$ & $0.025490450 - 0.0096494266 I$ & $0.0255404 - 0.00972708\,i$\\

		\end{tabular}
         \end{ruledtabular}
	\end{table*}

    \begin{table*}[t!]
    \caption{\label{tab:Comparingresults2} Comparing the quasinormal frequencies for the de Sitter modes ($n_{dS}=0$) for massless scalar field with $l=1$, from Ref.\cite{2020:Aragon} ($\omega_{dS}$), with our results obtained through Leaver ($\omega_{Leaver}$) and direct integration ($\omega_{DI}$).}
    \begin{ruledtabular}
    \begin{tabular}{|c|c|c|c|}

			$\Lambda M^{2}$ & $\omega_{dS} M$ \cite{2020:Aragon} & $\omega_{Leaver} M$ &$\omega_{DI}M$ \\ 
			\hline                               
			$0.02$ & $-0.081565496\,i$  & $-0.081565496\,i$ & $-0.0815655\,i$\\
			$0.04$ & $-0.11524810\,i$  & $-0.11524810\,i$ & $-0.115248\,i$\\
			$0.06$ & $-0.14100253\, i$  & $-0.14100253\,i$ & $-0.141003\,i$\\
			$0.08$ & $-0.16268011\,i$  & $-0.16268004\,i$ & $-0.16268\,i$\\
			$0.09$ & $-0.17249210\,i$  & $-0.17249210\,i$ & $-0.172492\,i$\\
			$0.1$ & $-0.18177480\,i$  & $-0.18177480\,i$ & $-0.181775\,i$\\
			$0.11$& $-0.19057630\,i$ & & $-0.1905 \,i$\\
			\hline

		\end{tabular}
         \end{ruledtabular}
	\end{table*}

\bibliography{MQNSdSrefs}

\end{document}